\documentclass[prl,superscriptaddress,aps]{revtex4}

\usepackage[T1]{fontenc}
\usepackage{setspace}
\usepackage{graphics}
\usepackage{amssymb}
\usepackage{color}

\begin{document}

\title{Neutron diffraction of hydrogenous materials: measuring incoherent and coherent intensities separately from liquid water - a 40-year-old puzzle solved}

%\title{Neutron diffraction of hydrogenous materials: measuring incoherent and coherent intensities separately over a momentum transfer range of 0.8-21~\AA$^{-1}$ from liquid water -- a 40-year-old puzzle solved}

%\title{Determination of the coherent and incoherent structure factors of mixtures of liquid heavy and light water over a momentum transfer range of 0.5-20~\AA$^{-1}$, using neutron diffraction with polarisation analysis}

\author{L\'aszl\'o Temleitner}
\affiliation{Institute for Solid State Physics and Optics, Wigner Research Centre for Physics, Hungarian Academy of Sciences, H-1121 Budapest, Konkoly Thege \'ut 29-33., Hungary}
\author{Anne Stunault}
\affiliation{Institute Laue Langevin, 71 Avenue des Martyrs, 38000 Grenoble, France}
\author{Gabriel Cuello}
\affiliation{Institute Laue Langevin, 71 Avenue des Martyrs, 38000 Grenoble, France}
\author{L\'aszl\'o Pusztai}
\affiliation{Institute for Solid State Physics and Optics, Wigner Research Centre for Physics, Hungarian Academy of Sciences, H-1121 Budapest, Konkoly Thege \'ut 29-33., Hungary}

\begin{abstract}

Accurate determination of the coherent static structure factor of any disordered material containing substantial amounts of proton nuclei has proven to be prohibitively difficult by neutron diffraction, due to the extremely large incoherent cross section of $^1$H. This notorious problem has continued to set severe obstacles to the reliable determination of liquid structures of hydrogenous materials up to this day, by introducing large uncertainties whenever a sample with a $^1$H content larger then about 20 \% had to be measured by neutron diffraction. Huge theoretical efforts over the past 40 years that had been aimed at estimating the 'incoherent background' of such data have not resulted in any practical solution to the problem. Here we present the first accurate separate measurements, by polarised neutron diffraction, of the coherent and incoherent contributions to the total static structure factor of mixtures of light and heavy water, over an unprecedentedly wide momentum transfer range, and over the 
entire composition range, i.e. for light water contents between 0 and 100 \%. We show that the measured incoherent background can be approximated by a Gaussian function. The separated coherent intensities exhibit signs of small inelastic contributions. Out of several possible approaches, we have chosen to subtract a cubic background using the Reverse Monte Carlo algorithm, which has the advantage of requiring an actual physical model (thousands of realistic water molecules at the correct density) behind the corrected data. Finally, coherent static structure factors for 5 different compositions of liquid H$_2$O and D$_2$O mixtures are presented for which the large incoherent background could actually be measured and separated, instead of being approximated as it has been done every time so far. These unprecedented experimental results provide a strong hope that determining the structure of hydrogenous materials, including, e.g., protein solutions, may become feasible in the near future.

\end{abstract}

\maketitle

\section*{}

Trustworthy information concerning the microscopic structure of liquid water is essential for most chemists and molecular biologists: life on Earth is based on water (and takes place in aqueous solutions). For this reason, liquid water has been the subject of a large number of diffraction studies (see, e.g., Refs.~ \cite{thiessen_82,soper_97,hura_00,hura_03,zeidler_11}). Computer simulation investigations using classical (for a review, see e.g. Ref.~\cite{hura_03}) and quantum mechanical (see e.g. Refs.~\cite{schwegler_00,wernet_04}) force fields abound. A key point here is that the development of force fields is - naturally - biased by experimental results; this is also a reason why reliable diffraction data are indispensable.

Despite voluminous relevant literature, liquid water is still viewed as one of the most notorious puzzles: for instance there are open questions concerning the average number of hydrogen bonds/molecule \cite{wernet_04, head-gordon_06}. It is also sometimes argued that an uncertainty exists regarding even the position of the first intermolecular O-H distance  -- a crucial one, as it characterises hydrogen bonding~\cite{pusztai_99}. The reason why these questions are still open is, quite clearly, the presence of \textit{hydrogen}.

X-ray diffraction is only little sensitive to hydrogen and can essentially only determine the oxygen-oxygen and, to some extent, oxygen-hydrogen correlations. In neutron diffraction the three distinct correlations of this binary system have a more similar weight: the difficulty is that for separating the three partial contributions, a contrast variation with $^1$H/$^2$H (or, in short, 'H/D') substitution is required. The $^1$H/$^2$H substitution, in principle, allows us to derive the most detailed information on the microscopic structure of hydrogenous (i.e., containing $^1$H) systems, due to nice contrast between their coherent neutron scattering lengths \cite{ncoherent} ($b_c^H=-3.7406fm$ and $b_c^D=6.671fm$ ). However, due to the exceptionally high level of spin-incoherence of $^1$H (spin-incoherent scattering cross-sections: $\sigma_i^H=80.27~barn$ and $\sigma_i^D=2.05~barn$), most of the measured diffraction signal from pure $^1$H$_2$O is incoherent background, useless from the structural point of 
view (see, e.g., Ref.~\cite{thiessen_82}). As a consequence, the structure factor of $^1$H$_2$O is still much debated \cite{thiessen_82,soper_97,pusztai_99}.

To make the situation even worse, in neutron diffraction, the structure factor of $^1$H$_2$O is the one with the highest information content, for the negative coherent scattering length of $^1$H: negative peaks would signify characteristic O-H distances in the total pair correlation function. Note that the $^1$H$_2$O structure factor is by far the most unreliable and therefore, even small errors in it have a large impact on the O-H partial pair correlation function. It has been clear for many decades that reliable neutron total structure factors of water samples with high light water content (ideally, of pure $^1$H$_2$O) would be decisive concerning H-bonding in water.
For this reason, numerous suggestions over the past 40 years have been made for the treatment of the huge 'incoherent background' (for an informative figure, see, e.g., \cite{thiessen_82}), none of which has proven to be routinely applicable (for the various approaches, see \cite{placzek_52,powles_72,blum_76,powles_79,granada_87a,dawidowski_94,palomino_07}). Real improvement could be expected only from an accurate experimental determination of the incoherent contributions from $^1$H.

Spin-incoherence, the major cause of the troubles, can in principle be tackled by separating the coherent and incoherent parts of the measured diffraction signal; this can be achieved by using polarised neutrons (see, e.g., Ref.~\cite{polar_textbook}). Interestingly, potentialities of polarised neutron diffraction have only been little exploited in this field; a possible reason for this is that available instruments provide data over only limited momentum transfer ranges, so that traditional evaluation, involving direct Fourier-transformation, would not be applicable. Concerning liquid water, only a couple of such investigations pop up \cite{dore_76,temleitner_07} in which coherent and incoherent intensities have been determined over a very limited momentum transfer range. Although these studies proved that the direct measurement of the incoherent background is, in principle, possible, the practical use of the results has remained marginal.

\begin{figure}[ht]
\begin{center}
\rotatebox{0}{\resizebox{0.4\textwidth}{!}{\includegraphics{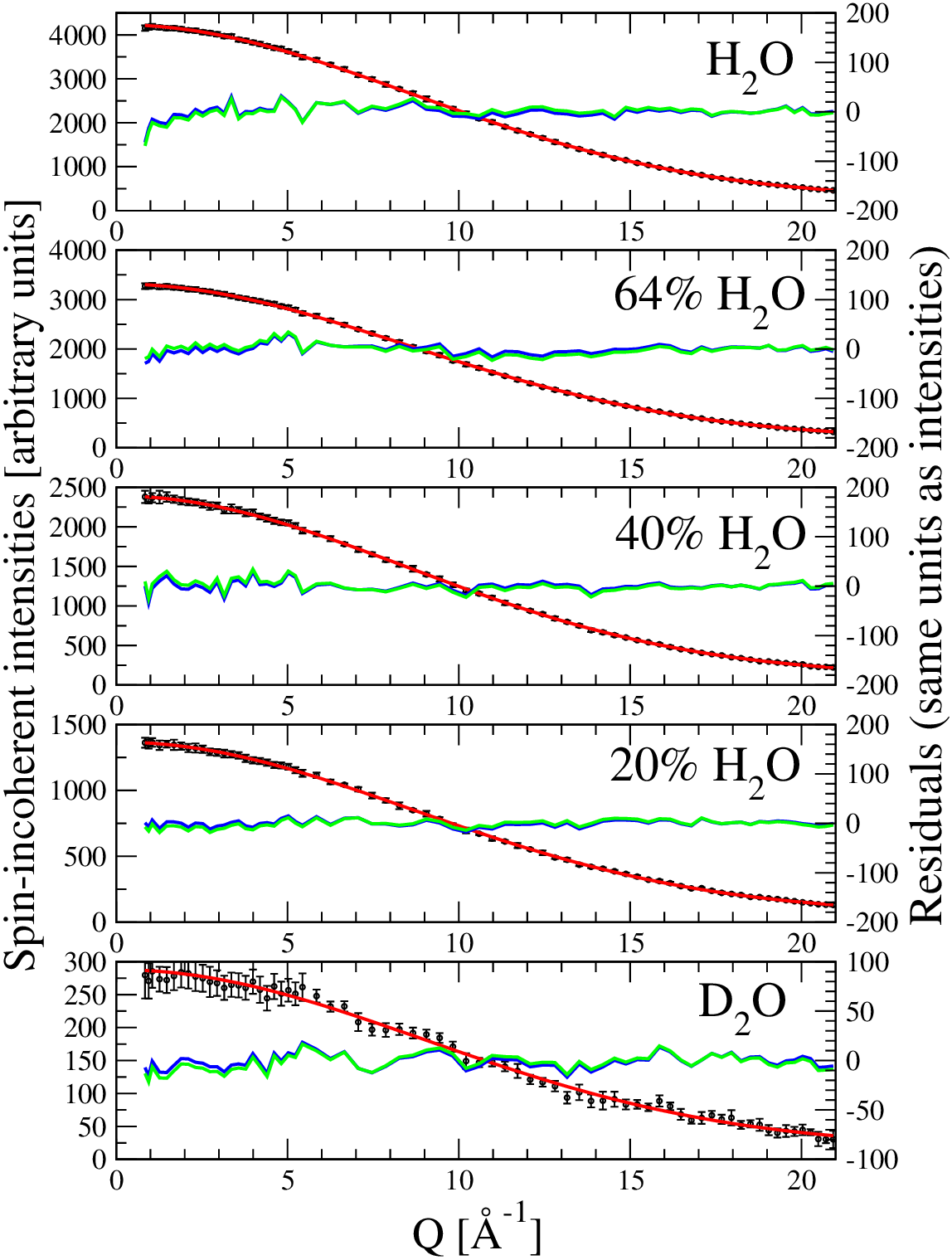}}}
\caption{\label{fig:incoh}
Incoherent intensities, obtained by multiplying the measured 'spin-flip' intensities by 3/2 (see eq. \ref{eq:incoh}), for 5 mixtures of light and heavy water. Light water content (from top to bottom): 100, 64, 40, 20 and 0 \%. Symbols with error bars: measurement; red solid lines: Gaussian (plus constant) fits; blue solid lines: residuals; green solid lines: residuals from the fit with a Gaussian of identical width. The statistical accuracy deteriorates as light water content decreases: this is just the manifestation of the well-known fact that pure heavy water (containing only $^2$H nuclei of the element hydrogen) shows only a small amount of spin incoherent scattering -- this is why neutron diffraction is most frequently performed on deuterated samples.%{All residuals magnified by 10x in the cases of other than 0\% light water content.}
}
\end{center}
\end{figure}

In the present investigation, the D3 instrument of the ILL (Grenoble, France) was used to measure the coherent and incoherent intensities from mixtures of light and heavy water over by far the widest ever Q-range reported for any hydrogenous material so far (for relevant details, see 'Methods'). Figure \ref{fig:incoh} shows the incoherent intensities, which are directly proportional to the spin incoherent cross sections, up 21~\AA$^{-1}$, along with numerical fits in the form of Gaussian functions plus a constant. Note that this functional form, that has not been suggested explicitly by any of the corresponding studies of the coherent background\cite{placzek_52,powles_72,blum_76,powles_79,granada_87a,dawidowski_94,palomino_07}, operates with only 3 adjustable parameters (see 'Methods'). The residuals are also shown in Figure \ref{fig:incoh}; it is obvious that the differences between fits and measured curves are exclusively of statistical nature. 
As shown by Table \ref{tab:incoh_parms}, the backgrounds for the various $^1$H contents can be parametrized by using one single Gaussian-width, with only a negligible worsening of the goodness-of-fit values.

\begin{figure}[ht]
\begin{center}
\rotatebox{0}{\resizebox{0.4\textwidth}{!}{\includegraphics{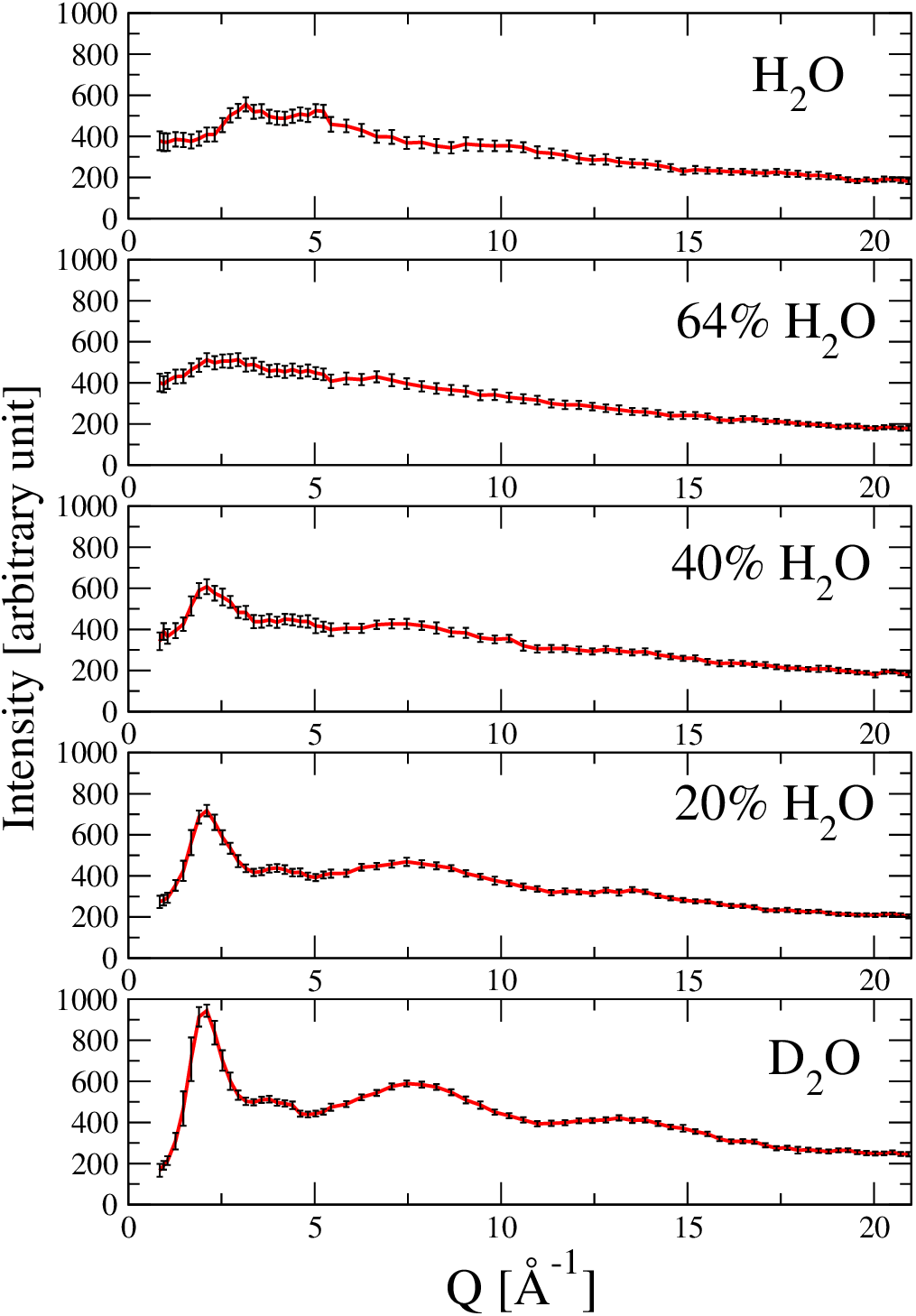}}}
\caption{\label{fig:coherent_measured}
Coherent intensities, obtained via eq. \ref{eq:coh}, for 5 mixtures of light and heavy water. Light water content (from top to bottom): 100, 64, 40, 20 and 0 \%.}
\end{center}
\end{figure}

In Figure \ref{fig:coherent_measured}, coherent intensities, as derived from a simple linear combination of the separately measured 'spin-flip' and 'non-spin-flip' intensities (see 'Methods'), are displayed.

Although the statistical accuracy may be further improved for the mixtures with the highest $^1$H content, it is already obvious that the functions are free from the enormous spin-incoherent background (cf. \cite{thiessen_82,dawidowski_12}). What is left to handle is a gentle droop towards higher Q-values that is characteristic to inelastic effects. Numerous approximations have been suggested to correct for such backgrounds, starting from the pioneering work of Placzek \cite{placzek_52} (which, as the author himself warned, is not really appropriate for light elements), to modern numerical fitting algorithms (see, e.g., \cite{CORRECT,hannon_90,soper_09,mcgr_97,rmcpp}). 

\begin{figure}[ht]
\begin{center}
\rotatebox{0}{\resizebox{0.4\textwidth}{!}{\includegraphics{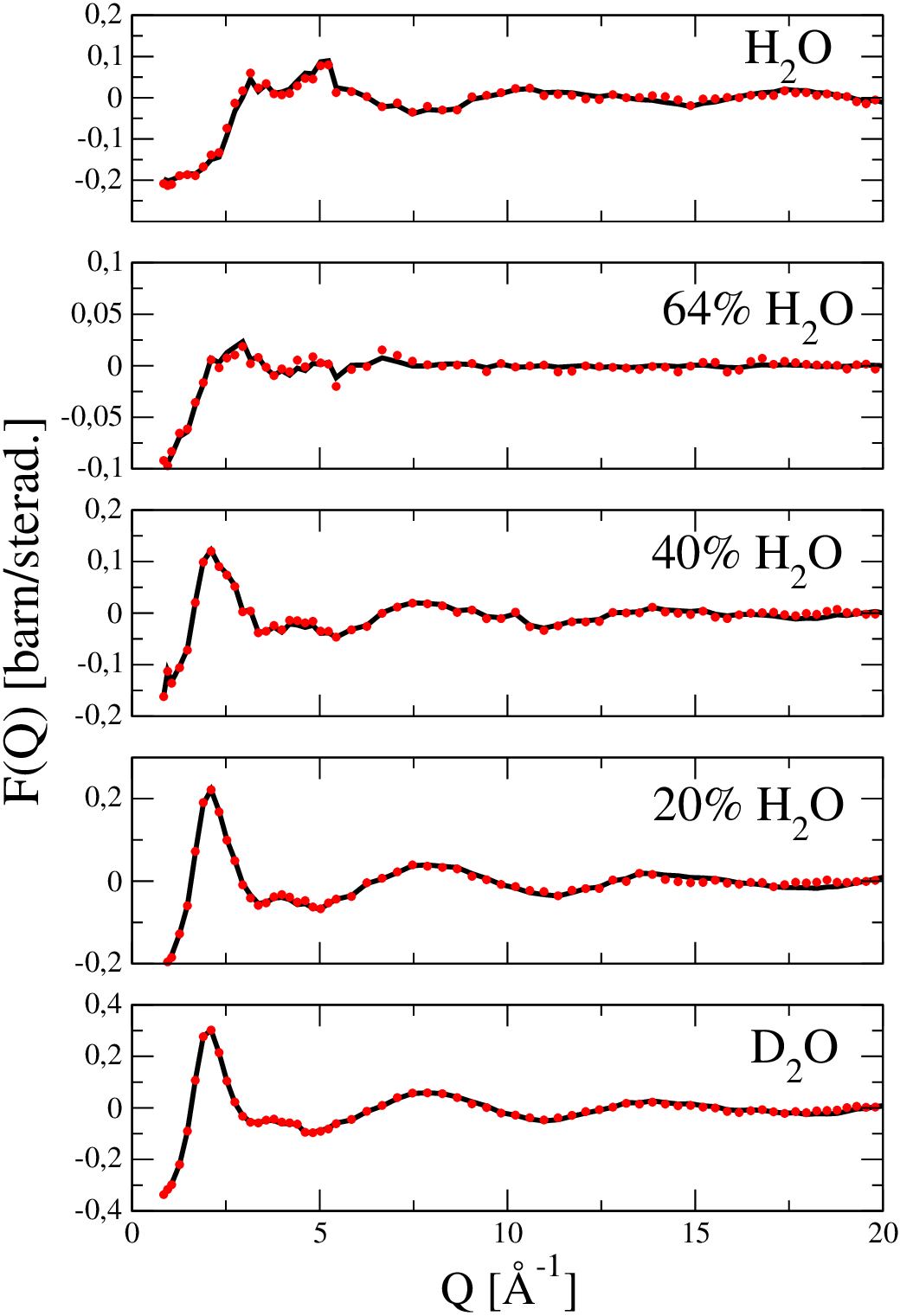}}}
\caption{\label{fig:coherent_rmc}
Coherent static structure factors for 5 mixtures of light and heavy water. Light water content (from top to bottom): 100, 64, 40, 20 and 0 \%. Red dots: experimental points; black solid lines: Reverse Monte Carlo fits.}
\end{center}
\end{figure}

For our present purposes, the Reverse Monte Carlo (RMC) method of structural modelling has been selected (for more details, see 'Methods'). The main reason for this choice was that in RMC, thousands of realistic molecules have to be present at the correct density in physically existing simulation boxes; that is, the limiting values of the radial distribution functions and structure factors are automatically built-in. Then, a cubic background may also be refined, simultaneously with the usual approach to the measured data via random moves of the particles (see 'Methods'). Results from such RMC calculations are presented in Figure \ref{fig:coherent_rmc} for all five mixtures of light and heavy water, in the form of normalised coherent static structure factors. Again, it may be argued that, statistically, measured curves for the mixtures with the highest $^1$H contents are not as good as the ones that correspond to the $^2$H-dominated samples. It is nevertheless clear that corrections for inelasticity could 
also be performed successfully. That is, the measurement-based determination of the spin-incoherent background for hydrogenous samples has directly led to coherent static structure factors, a result that has been awaited for decades.

\begin{figure}[ht]
\begin{center}
\rotatebox{0}{\resizebox{1.0\textwidth}{!}{\includegraphics{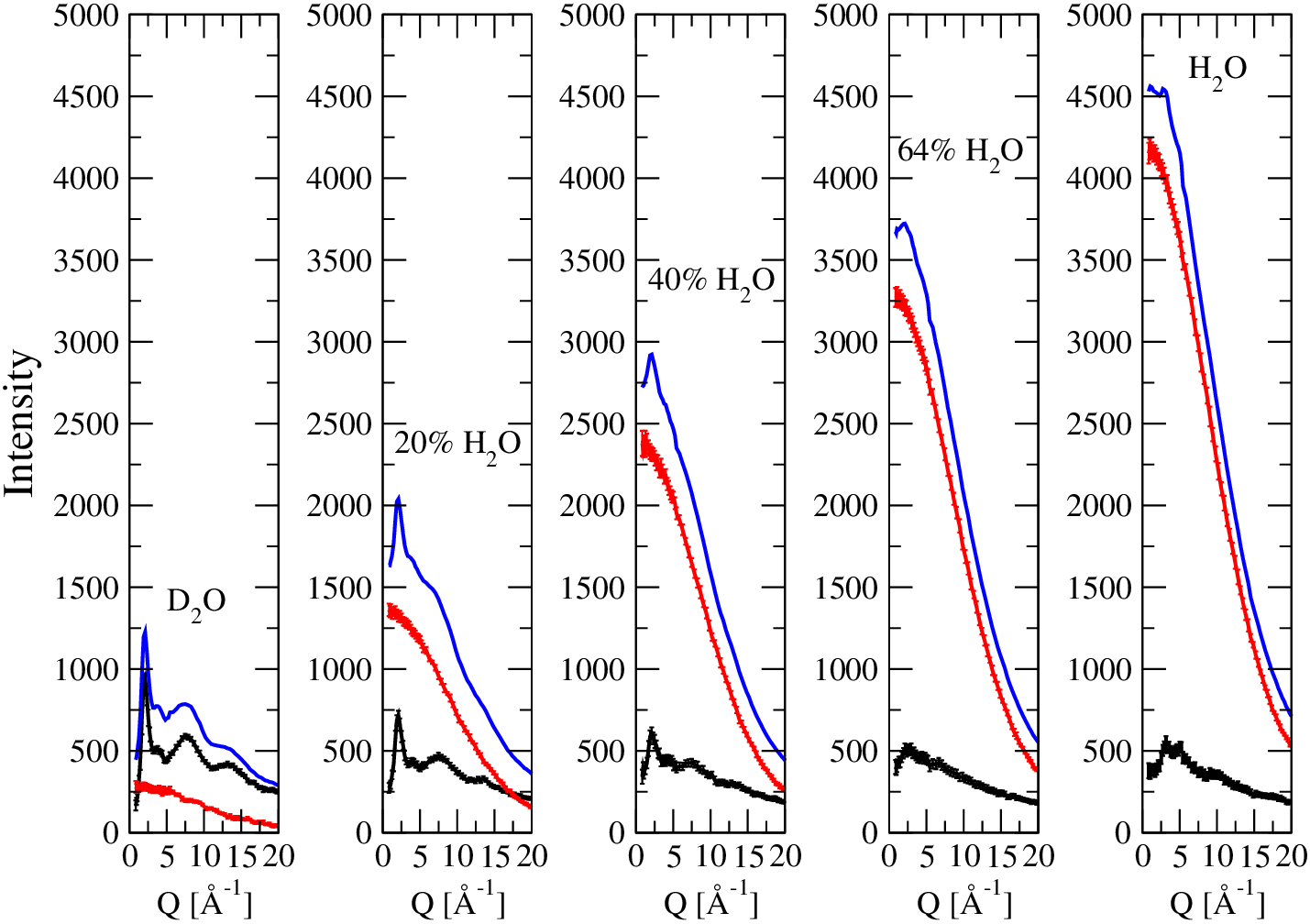}}}
\caption{\label{fig:polviz_demo}
Measured spin-incoherent (red lines) and coherent (black lines) intensities for 5 mixtures of light and heavy water. Light water content (from left to right): 0, 20, 40, 64 and 100 \%. Blue lines: sums of the (here, separately measured) coherent and incoherent intensities, which therefore represent the neutron diffraction signal measurable without polarization analysis. If one wishes to reach the coherent intensities (black curves) from non-polarised data then a large number must be approximated (individual points of the red curves) and subtracted from another large number (individual points of the blue curves) and the desired result is a small number (individual points of the black curves). That is, taking the standard (non-polarised) way, the statistical errors only are large enough to render the entire analysis problematic, not to mention systematic uncertainties in conjunction with estimating the (spin-)incoherent contributions. 
}
\end{center}
\end{figure}

For a demonstration of the achievements, in Figure \ref{fig:polviz_demo} the separated coherent and incoherent intensities are displayed, together with their sums; the sums represent the information measurable with standard (i.e., non-polarised) neutron diffraction. It is clear that even in the case of pure heavy water, the precise knowledge of the incoherent intensity is desirable. For samples with more than (about) 20 \% $^1$H-content, the incoherent contributions are much greater than the coherent ones and therefore, proper handling of them is only possible if one can measure both contributions separately; even then, gathering sufficiently good statistics may be an issue.

Summarising the achievements reported here, (1) the incoherent and coherent scattered intensities from hydrogenous water samples, including pure light water, have been measured by polarised neutron diffraction over an unprecedentedly wide momentum transfer range; (2) the incoherent intensities could be described, within statistical errors, by Gaussian curves and a constant, using three adjustable parameters only, which finding may later be used for correcting data from non-polarised neutron diffraction; (3) the separated coherent intensities are shown to contain roughly equal proportions of inelastic contributions, which is consistent with the notion that their origin lies in molecular recoil effects (as opposed to single-atom relaxation); (4) this remaining inelastic background could easily be corrected for by using the Reverse Monte Carlo method of structural modelling, thus providing the coherent static factors for hydrogenous materials with an arbitrary amount of protons $^1$H. 

There are far reaching consequences of the present findings: (1) it has now been demonstrated that hydrogenous samples without deuteration can be handled by (polarised) neutron diffraction, which, in turn, means that the microscopic structure of 'soft matter' systems of any complexity can be tackled; (2) the H/D isotopic substitution, wherever feasible, may now gain an enhanced reputation, since the large contrast between the coherent scattering lengths of $^1$H and $^2$H may be fully exploited; (3) it is now clear that for a proper consideration of incoherent contributions from hydrogenous samples one needs to measure them at least up to Q-values as reported here and the angular range should be as high as possible, otherwise the points of inflection (between 10 and 15~\AA$^{-1}$, see Figure \ref{fig:incoh}) will be missed; (4) since this takes nearly prohibitively long experiments (ideally, as we see it now, of the order of one full week on the D3 instrument for just one hydrogenous sample 
with $^1$H only), a very carefully planned list of the 'reference' samples, to be measured by any means, has to be drawn;
(5) we envisage an inevitable (although, perhaps only moderate) boost in developing polarised neutron instrumentation capable of the experiments described here; (6) emphasis may still have to be put on the more and more precise description of the incoherent background (cf. Figure \ref{fig:incoh}), so that the functional form could be applied for correcting data taken at neutron diffractometers without polarisation analysis (as opposed to measuring more and more samples  -- there may not be enough beamtime for practising both kinds of activities freely). 

\section{Methods}
\subsection{Neutron diffraction with polarisation analysis}

Diffraction experiments using polarised neutrons have been conducted on the D3 instrument \cite{D3-REFERENCE} installed on the hot source of the Institut Laue-Langevin (ILL; Grenoble, France). Five isotopic compositions of $^1$H (protons) and $^2$H (deuterons) have been prepared, containing 0, 20, 40, 64 ('zero water', see e.g. \cite{temleitner_07}), and 100 \% light water (with $^1$H). Liquid samples were put in a double-walled vanadium container (internal diameter: 8 mm, outer diameter: 10.7 mm), in order to minimise contributions from multiple scattering; the sample geometry was therefore that of a hollow cylinder.  The experiments were performed at ambient pressure and temperature. Using the D3 instrument with 0.5~\AA{} wavelength neutrons, scattering intensities have been collected in both spin-flip and non-spin-flip modes over a uniquely wide momentum transfer range of 0.8-21~\AA$^{-1}$ (4-120 degrees in $2\Theta$). This outstanding coverage of the reciprocal space can be achieved by making use of a 
Heussler-alloy polariser and a $^3$He analyser cell that contains spin-polarised nuclei \cite{surkau_97}. Samples with the highest $^1$H content have been investigated for somewhat longer time than those dominantly with $^2$H, so that statistics of the coherent signals would be comparable. Still, the measuring time of about 24 hours for pure light water provided statistics somewhat poorer than hoped for. The usual corrections\cite{heil_02} for polarisation efficiency have also been carried out before further data processing. Coherent and spin-incoherent contributions to the total scattering have been separated in the usual manner~\cite{polar_textbook}, using the following formulae:

\begin{equation}\label{eq:coh}
I_{coh}(Q)=I^{NSF}(Q) - \frac{1}{2} I^{SF}(Q)
\end{equation}

and

\begin{equation}\label{eq:incoh}
I_{incoh}(Q) = \frac{3}{2} I^{SF}(Q)
\end{equation}

where the "NSF" and "SF" indices refer to intensities measured in "non-spin-flip"
and "spin-flip" modes, respectively. 

\subsection{Fitting the incoherent intensity curves}

Following a couple of trial attempts, it has become clear that a single Gaussian and an additional constant are perfect for fitting the measured signals within errors (see Figure \ref{fig:incoh}).  The functional form is then:
\begin{equation}
 I_{incoh}(Q) = Intensity\frac{2}{FWHM}\sqrt{\frac{\ln (2)}{\pi}}*\exp\left\{-4\ln (2) \left( \frac{Q}{FWHM}\right)^2\right\} + Constant
\end{equation}

There are only three parameters to adjust here, {\it Intensity}, {\it FWHM} and {\it Constant}; these are listed in Table \ref{tab:incoh_parms} for the 5 mixtures of light and heavy water considered in this study. Fitting for each incoherent dataset was performed for 71 data ({\it Q}) points, using our own computer program that implements a non-linear least-squares fitting algorithm by the steepest descent method. Furthermore, the similarity of the Gaussian width parameters obtained for the individual measurements suggested that a fit using a common (identical) width parameter for all measurements might be applicable. The identical width of $FWHM=20.128(25)\textrm{~\AA}^{-1}$ provided only slightly worse  goodness-of-fit (R$_{wp}$, as defined in \cite{rietveld_99}) values than the individual parametrization, so that agreements between fits and original curves are still good.
\begin{table}[ht]
\begin{center}
\footnotesize
\begin{tabular}{||c||c|c|c|c||c|c|c||}
\hline \hline
$^1$H$_2$O & \multicolumn{4}{c||}{Individual fits} & \multicolumn{3}{c||}{Fit with common FWHM} \\
\cline{2-8}
content [\%]  & Intensity & FWHM [\AA$^{-1}$] & Constant term & R$_{wp}$ [\%] & Intensity & Constant term &  R$_{wp}$ [\%] \\
\hline \hline
100 & 85550(163) & 20.217(31) & 258.8(2.9) & 0.53 &  85140(166) & 266.7(2.7) & 0.56 \\
\hline
64 & 66752(120) & 19.974(30) & 175.0(2.1) & 0.58 & 67267(135) & 165.7(2.12) & 0.68 \\
\hline
40 & 48414(114) & 19.955(40) & 111.8(2.2) & 0.77 & 48814(113) & 102.8(1.8) & 0.87 \\
\hline
20 & 28323(102) & 20.381(61) & 61.6(2.1) & 1.08 & 27980(68) & 69.5(1.16) & 1.21 \\
\hline
0 & 6093(146) & 21.193(385) & 17.8(3.2) & 4.75 & 5746(45) & 25.8(0.88) & 5.04 \\
\hline \hline
\end{tabular}
\caption{\label{tab:incoh_parms} 
Adjusted parameters {\it Intensity}, {\it FWHM} and {\it Constant term}, together with goodness-of-fit (R$_{wp}$) values while fitting the measured incoherent intensities with individual and with common FWHM ($20.128(25)\textrm{~\AA}^{-1}$) parameters. (The number of degrees of freedom was 68 for each individual fits and 344 for the common FWHM fit during the actual calculations.)}
\end{center}
\end{table}

\subsection{Reverse Monte Carlo modeling}

Reverse Monte Carlo (RMC)\cite{rmc,rmcpp} is a tool for constructing large, three-dimensional structural models that are consistent, within the estimated level of experimental errors, with measured total scattering structure factors obtained from diffraction experiments. Via random movements of atoms, the difference between experimental and model total structure factors is minimized. As a result, by the end of the calculation a particle configuration is available that is consistent with the experimental structure factor(s). If the structure is analysed further, partial radial distribution functions, as well as other structural characteristics (neighbour distributions, cosine distribution of bond angles) can be calculated from the particle configurations. In the present study, however, we only wished to exploit the following useful features of the RMC algorithm:  (1) the resulting total scattering structure factors, due to the requirement that they must belong to large physical (simulated) particle 
configurations, must adhere to important limiting values; and (2) they also have to be consistent with basic characteristics (such as density, particle sizes, and the molecular nature) of the system.

In order to remove the remaining (isotope) incoherent background from the experimental coherent scattering intensities and normalise the results to the same scale, reverse Monte Carlo modelling calculations have been performed with 2000 molecules, using an atomic number density of $0.10028$~\AA$^{-3}$. The $H_2O$ unit was kept together by using fixed neighbour constraints \cite{rmcpp}, allowing small variances of the interatomic OH ($0.98\pm 0.02$~\AA) and HH ($1.55\pm 0.03$~\AA) distances. Between atoms belonging to different molecules, closest approach distances were applied (OO: 2.2~\AA, OH: 1.5~\AA, HH: 1.6~\AA).

The correction terms that have been subtracted as background have the form:
\begin{equation}
 F_{coh}(Q)=r*F_{measured}(Q)-\left(a+bQ+cQ^2+dQ^3\right)
\end{equation}
where parameters {\it r, a, b, c} and {\it d} are provided in Table \ref{tab:rmc_parms}.

\begin{table}[ht]
\begin{center}
\small
\begin{tabular}{||c||c|c|c|c|c||}
\hline \hline
$^1$H$_2$O content [\%]  & r [x$10^{-3}$] & a [x$10^{-1}$] & b [x$10^{-3}$] & c [x$10^{-3}$] & d [x$10^{-5}$] \\
\hline \hline
100 & 1.15 & 6.66 & -26.2 & -0.474 &  3.37 \\
\hline
64 & 0.744 & 4.02 & -11.9 & -0.595 &  2.65 \\
\hline
40 & 1.01 & 5.11 & -4.40 & -1.50 &  4.60 \\
\hline
20 & 0.963 & 4.70 & 2.68 & -1.77 &  4.80 \\
\hline
0 & 0.819 & 4.72 & 2.98 & -1.48 &  3.25 \\
\hline \hline
\end{tabular}
\caption{\label{tab:rmc_parms}  Parameters of the subtracted background from the coherent intensities.}
\end{center}
\end{table}

\section*{Acknowledgment}
The authors thank Dr. Werner Schweika (FZ J\"ulich, Germany) for his valuable advices and help provided during the initial stage of the experimental work. LT and LP wish to thank the Hungarian Basic Research Fund (OTKA) for financial support via grant No. K083529. Beamtime on the D3 instrument provided by the Institut Laue Langevin (Grenoble, France), under proposal numbers 6-02-362, 6-02-519 and TEST-2089, is gratefully acknowledged. We thank Ms. A. Szuja (Research Centre for Energy, Hungarian Academy of Sciences) for carefully preparing all mixtures of light and heavy water. 

\section*{Author contributions}
LT, AS, GJC and LP have performed the experiments; AS and GJC have modified the D3 instrument for liquid samples; AS has performed initial data analyses; LT has coded software for data processing and for the Gaussian fitting; LT has performed RMC modeling; LP wrote the first version of the manuscript. All authors discussed the results and commented on the manuscript.

\section*{References}

\bibliographystyle{unsrt}
\bibliography{polwater}

\begin{thebibliography}{10}

\bibitem{thiessen_82}
W.~E. Thiessen and A.~H. Narten.
\newblock {Neutron diffraction study of light and heavy water mixtures at
  25$^o$C}.
\newblock {\em Journal of Chemical Physics}, 77:2656--2662, 1982.

\bibitem{soper_97}
A.~K. Soper, F.~Bruni, and M.~A. Ricci.
\newblock {Site--site pair correlation functions of water from 25 to 400$^o$C:
  Revised analysis of new and old diffraction data}.
\newblock {\em Journal of Chemical Physics}, 106:247--254, 1997.

\bibitem{hura_00}
G.~Hura, J.~M. Sorenson, R.~M. Glaeser, and T.~Head-Gordon.
\newblock {A high-quality x-ray scattering experiment on liquid water at
  ambient conditions}.
\newblock {\em Journal of Chemical Physics}, 113:9140--9148, 2000.

\bibitem{hura_03}
G.~Hura, D.~Russo, R.~M. Glaeser, T.~Head-Gordon, M.~Krack, and M.~Parinello.
\newblock {Water structure as a function of temperature from X-ray scattering
  experiments and ab initio molecular dynamics}.
\newblock {\em Physical Chemistry Chemical Physics}, 5:1981--1991, 2003.

\bibitem{zeidler_11}
A.~Zeidler, P.~S. Salmon, H.~E. Fischer, J.~C. Neuefeind, J.~M. Simonson,
  H.~Lemmel, H.~Rauch, and T.~E. Markland.
\newblock {Oxygen as a Site Specific Probe of the Structure of Water and Oxide
  Materials}.
\newblock {\em Physical Review Letters}, 107:145501, 2011.

\bibitem{schwegler_00}
E.~Schwegler, G.~Galli, and F.~Gygi.
\newblock {Water under Pressure}.
\newblock {\em Physical Review Letters}, 84:2429--2432, 2000.

\bibitem{wernet_04}
P.~Wernet, D.~Nordlund, U.~Bergmann, M.~Cavallieri, M.~Odelius, H.~Ogasawara,
  L.~{\AA}. H{\"a}slund, T.~K. Hirsch, L.~Ojem{\"a}e, P.~Glatzel, L.~G.~M.
  Pettersson, and A.~Nilsson.
\newblock {The Structure of the First Coordination Shell in Liquid Water}.
\newblock {\em Science}, 304:995--999, 2004.

\bibitem{head-gordon_06}
T.~Head-Gordon and M.~E. Johnson.
\newblock {Tetrahedral structure or chains for liquid water}.
\newblock {\em Proceedings of the National Academy of Sciences of the United
  States of America}, 103:7973--7977, 2006.

\bibitem{pusztai_99}
L.~Pusztai.
\newblock {Partial pair correlation functions of liquid water}.
\newblock {\em Physical Review B}, 60:11851--11854, 1999.

\bibitem{ncoherent}
V.~F. Sears.
\newblock {Neutron scattering lengths and cross sections}.
\newblock {\em Neutron News}, 3(3):26--37, 1992.

\bibitem{placzek_52}
G.~Placzek.
\newblock {The Scattering of Neutrons by Systems of Heavy Nuclei}.
\newblock {\em Physical Review}, 86:377--388, 1952.

\bibitem{powles_72}
J.~G. Powles, J.~C. Dore, and D.~I. Page.
\newblock {Coherent neutron scattering by light water ($H_2O$) and a
  light-heavy water mixture (64 per cent $H_2O$/36 per cent $D_2O$)}.
\newblock {\em Molecular Physics}, 24:1025--1037, 1972.

\bibitem{blum_76}
L.~Blum and A.~H. Narten.
\newblock {Diffraction by Molecular Liquids}.
\newblock In I.~Prigogine and S.~A. Rice, editors, {\em {Advances in Chemical
  Physics}}, volume~34, pages 203--243. John Wiley \& Sons, Inc., 1976.

\bibitem{powles_79}
J.~G. Powles.
\newblock {Slow neutron scattering by molecules V. Recoil corrections for any
  molecule}.
\newblock {\em Molecular Physics}, 37:623--641, 1979.

\bibitem{granada_87a}
J.~R. Granada, V.~H. Gillette, and R.~E. Mayer.
\newblock {Calculation of neutron cross sections and thermalization parameters
  for molecular gases using a synthetic scattering function. I}.
\newblock {\em Physical Review A}, 36:5585--5593, 1987.

\bibitem{dawidowski_94}
J.~Dawidowski, J.~R. Granada, R.~E. Mayer, G.~J. Cuello, V.~H. Gillette, and
  M.-C. Bellissent-Funel.
\newblock {Multiple scattering and ineasticity corrections in thermal neutron
  scattering experiments on molecular systems}.
\newblock {\em Physica B}, 203:116--128, 1994.

\bibitem{palomino_07}
L.~A. {Rodr{\'i}gez Palomino}, J.~Dawidowski, J.~J. Blostein, and G.~J. Cuello.
\newblock {Data processing method for neutron diffraction experiments}.
\newblock {\em Nuclear Instruments and Methods in Physics Research B},
  258:453--470, 2007.

\bibitem{polar_textbook}
Th. Br{\"u}ckel and W.~Schweika, editors.
\newblock {\em {Polarized Neutron Scattering}}, volume~12 of {\em {Schriften
  des Forschungszentrums J{\"u}lich, Reihe Materie und Material/Matter and
  Materials}}.
\newblock Forschungszentrum J{\"u}lich GmbH, 2002.

\bibitem{dore_76}
J.~C. Dore, J.~H. Clarke, and J.~T. Wenzel.
\newblock {Separation of coherent and spin-incoherent neutron scattering by
  polarization analysis}.
\newblock {\em Nuclear Instruments and Methods}, 138:317--319, 1976.

\bibitem{temleitner_07}
L.~Temleitner, L.~Pusztai, and W.~Schweika.
\newblock {The structure of liquid water by polarized neutron diffraction and
  reverse Monte Carlo modelling}.
\newblock {\em Journal of Physics: Condensed Matter}, 19:335207, 2007.

\bibitem{dawidowski_12}
J.~Dawidowski and G.~J. Cuello.
\newblock {Experimental corrections in neutron diffraction of ambient water
  using H/D isotopic substitution}.
\newblock {\em Journal of Physics: Conference Series}, 340:012004, 2012.

\bibitem{CORRECT}
M.~A. Howe, R.~L. McGreevy, and W.~S. Howells.
\newblock {The analysis of liquid structure data from time-of-flight neutron
  diffractometry}.
\newblock {\em Journal of Physics: Condensed Matter}, 1:3433--3451, 1989.

\bibitem{hannon_90}
A.~C. Hannon, W.~S. Howells, and A.~K. Soper.
\newblock {\em Intitute of Physics Conference Series}, 107:193, 1990.

\bibitem{soper_09}
A.~K. Soper.
\newblock {Inelasticity corrections for time-of-flight and fixed wavelength
  neutron diffraction experiments}.
\newblock {\em Molecular Physics}, 107:1667--1684, 2009.

\bibitem{mcgr_97}
L.~Pusztai and R.~L. McGreevy.
\newblock {MCGR: An inverse method for deriving the pair correlation function
  from the structure factor}.
\newblock {\em Physica B}, 234-236:357--358, 1997.

\bibitem{rmcpp}
O.~Gereben, P.~J{\'o}v{\'a}ri, L.~Temleitner, and L.~Pusztai.
\newblock {A new version of the RMC++ Reverse Monte Carlo programme, aimed at
  investigating the structure of covalent glasses}.
\newblock {\em Journal of Optoelectronics and Advanced Materials},
  9:3021--3027, 2007.

\bibitem{D3-REFERENCE}
E.~Leli{\`e}vre-Berna, E.~Bourgeat-Lami, Y.~Gilbert, N.~Kernavanois,
  J.~Locatelli, T.~Mary, G.~Pastorello, A.~Petukhov, S.~Pujol, R.~Rouques,
  F.~Thomas, M.~Thomas, and F.~Tasset.
\newblock {ILL polarised hot-neutron beam facility D3}.
\newblock {\em Physica B}, 356:141--145, 2005.

\bibitem{surkau_97}
R.~Surkau, J.~Becker, M.~Ebert, T.~Grossmann, W.~Heil, D.~Hofmann, H.~Humblot,
  M.~Leduc, E.~W. Otten, D.~Rohe, K.~Siemensmeyer, M.~Steiner, F.~Tasset, and
  N.~Trautmann.
\newblock {Realization of a broad band neutron spin filter with compressed,
  polarized $^3$He gas}.
\newblock {\em Nuclear Instruments and Methods in Physics Research A},
  384:444--450, 1997.

\bibitem{heil_02}
W.~Heil, K.~H. Andersen, R.~Cywinski, H.~Humblot, C.~Ritter, T.~W. Roberts, and
  J.~R. Stewart.
\newblock {Large solid-angle polarisation analysis at thermal neutron
  wavelengths using a $^3$He spin filter}.
\newblock {\em Nuclear Instruments and Methods in Physics Research A},
  485:551--570, 2002.

\bibitem{rietveld_99}
L.~B. McCusker, R.~B. {Von Dreele}, D.~E. Cox, D.~Lou{\"e}r, and P.~Scardi.
\newblock {Rietveld refinement guidelines}.
\newblock {\em Journal of Applied Crystallography}, 32:36--50, 1999.

\bibitem{rmc}
R.~L. McGreevy and L.~Pusztai.
\newblock {Reverse Monte Carlo Simulation: A New Technique for the
  Determination of Disordered Structures}.
\newblock {\em Molecular Simulation}, 1:359--367, 1988.

\end{thebibliography}

\end{document}